\documentclass[aps,prd,preprint,nofootinbib]{revtex4-2}
\pdfoutput=1

\usepackage[T1]{fontenc}
\usepackage[utf8]{inputenc}
\usepackage{graphicx}
\usepackage{amsmath}
\usepackage{amssymb}
\usepackage{siunitx}
\usepackage{booktabs}
\usepackage{longtable}
\usepackage{hyperref}
\graphicspath{{figs/}}
\begin{document}
\title{Agnostic search for a vector-like top partner at a lepton collider via the recoil mass technique}

\author{Higinio Valle}
\author{Cristina Oropeza Barrera}
\email{cristina.oropeza@ibero.mx}
\author{Oscar Ochoa-Valeriano}
\email{oscar.ochoa01@correo.uia.mx}
\affiliation{Departamento de Física y Matemáticas, Universidad Iberoamericana Ciudad de México, Prol. Paseo de la Reforma 880, Lomas de Santa Fe, Mexico City 01219, Mexico}

\date{\today}

\begin{abstract}
A study of the single production of a vector-like top partner $T$ at a $\sqrt{s}=3$~TeV $e^{+}e^{-}$ collider is presented, using a decay-mode and mass-agnostic event selection built around the recoil mass observable. The analysis relies on fully simulated samples processed through a common reconstruction and event selection chain. The baseline selection is validated without beam polarization or initial-state radiation (ISR). The impact of more realistic running conditions, including ISR together with an $e^{-}$ beam polarization of $-80\%$, is then evaluated and the discovery reach is mapped as a function of the top partner mass $m_{T}$ and coupling strength $\kappa_{T}$. In the baseline scan, the estimated statistical significance rises above $3\sigma$ for the lowest benchmark masses once $\kappa_{T}$ exceeds approximately 0.3.
\end{abstract}


\maketitle
\section{Introduction}

Vector-like quarks (VLQs) are among the best-motivated extensions of the Standard Model (SM) in the fermionic sector, particularly in scenarios that address the Hierarchy Problem \cite{hierarchy, peskin2025hierarchyproblem} through new dynamics in the top sector. They arise naturally in Composite Higgs constructions~\cite{composite_vector_resonances, composite-Higgs_minimal, composite-Higgs_phenomenology}, Little-Higgs realizations~\cite{little-Higgs}, and extra-dimensional frameworks~\cite{Antoniadis:1990}. In general, VLQs are introduced in 4 flavours: $T$ (+2/3), $B$ (-1/3), $X$ (+5/3), $Y$ (-4/3), and are assumed to couple primarily to the third generation of SM quarks and electro-weak (EW) bosons \cite{VLQ_handbook}. Gauge-invariant mass terms appear in the Lagrangian since their non-chiral nature arises from the lack of Yukawa couplings to the Higgs field-- as such, precision measurements in the EW sector do not set bounds on their couplings to SM quarks \cite{Eberhardt_2012}. The decay modes and branching fractions for these hypothetical particles depend on the chosen multiplet representation of the weak isospin, $SU(2)_L$, as well as on the peculiarities of the non-minimal models.
 
Current constraints on VLQs come from searches at the Large Hadron Collider (LHC) \cite{CMS_review, ATLAS_review, cmscollaboration2026searchsingleproductionvectorlike}, where large QCD rates are accompanied by substantial combinatorial and irreducible backgrounds. These searches are typically optimized for specific decay hypotheses, so their sensitivity can deteriorate rapidly away from the assumed branching-fraction pattern. A multi-TeV $e^{+}e^{-}$ collider offers a complementary and cleaner environment, particularly for single-production channels in which the sensitivity is directly affected by the size of the effective coupling $\kappa$. The proposed CLIC program \cite{brunner2022clicproject} at $\sqrt{s}=3$~TeV is well suited to explore this regime. Recent phenomenological studies have focused on searches for single- and pair-production of $T$ \cite{TCLICsearch, Qin:2023CLIC, Yang:2023CLIC} and $B$ \cite{BCLICsearch, benbrik2026sensitivitysinglevectorlikequark, Yang_2026} in specific decay modes.

In this work, the process $e^{+}e^{-}\to T t$ is studied in a model setup following Ref.~\cite{Buchkremer:2013}. The analysis strategy is intentionally agnostic to both the top-partner mass $m_T$ and its decay mode. Unlike traditional LHC searches, which often rely on narrow resonance windows and channel-specific selections, the present approach exploits the known initial state of a lepton collider through the recoil mass against a reconstructed hadronic SM top quark candidate. This observable preserves sensitivity across a broad range of branching-fraction hypotheses without tailoring the selection to a single topology. Baseline sensitivities are first established using samples without initial-state radiation (ISR) or beam polarization in order to calibrate the intrinsic performance of the recoil-based selection. The impact of more realistic running conditions, including ISR together with $P_{e^-}=-80\%$, is then quantified. Finally, the expected significance in the $(m_T,\kappa_T)$ plane is mapped to provide a compact interpretation of the discovery reach.

\section{Signal and background samples}

This analysis uses the vector-like quark model developed by Buchkremer et al. \cite{Buchkremer:2013}. The top partner is introduced as a singlet that mixes exclusively with SM third-generation quarks and decays to $bW$ (50\%), $tZ$ (25\%) and $tH$ (25\%)\footnote{The Goldstone equivalence theorem \cite{Cornwall:1974km} yields the (normalized) decay rate to EW bosons in the asymptotic limit where the mass $M$ of the VLQ goes to infinity.}. The effective Lagrangian description is given by,
\begin{align*}
  \mathcal{L}_{\mathrm{eff}} = \kappa_T \Biggl\{
  & \sqrt{\frac{\zeta \xi_W^T}{\Gamma_W^0}}\,\frac{g}{\sqrt{2}}
    \left[\Bar{T}_{L/R} W_\mu^+ \gamma^{\mu} b_{L/R}\right] \\
  & + \sqrt{\frac{\zeta \xi_Z^T}{\Gamma_Z^0}}\,\frac{g}{2c_W}
    \left[\Bar{T}_{L/R} Z_\mu \gamma^{\mu} t_{L/R}\right] \\
  & - \sqrt{\frac{\zeta \xi_H^T}{\Gamma_H^0}}\,\frac{M}{v}
    \left[\Bar{T}_{R/L} H t_{L/R}\right]
  \Biggr\} + \text{h.c.}
\end{align*}

\begin{figure}[!ht]
  \centering
  \includegraphics[width=0.75\textwidth, height=5cm, trim={1cm 21.5cm 2cm 1cm}, clip]{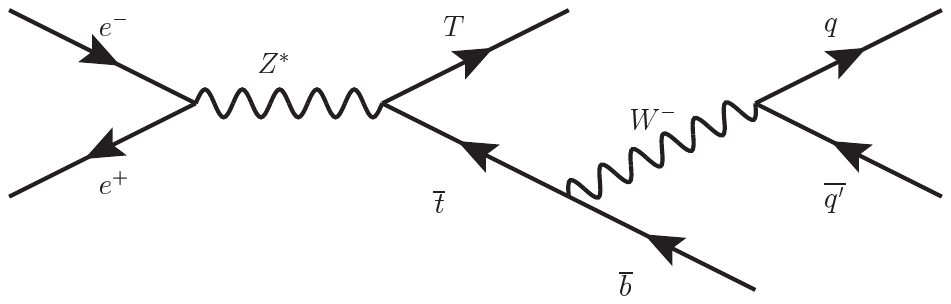}
  \caption{Example Feynman diagram of the signal process.}
  \label{fig:feynman}
\end{figure}

The signal process explored in this study, $e^{+}e^{-}\to T\,t$, is shown in Figure~\ref{fig:feynman}. Given the high masses of the top partners, the projected lepton collider that is the most suitable for this analysis is CLIC \cite{brunner2022clicproject} as it is the only one that could reach the necessary center-of-mass energies to produce such heavy particles. Therefore, the run conditions of CLIC-3 are used, including $\sqrt{s}=3$~TeV, electron-beam polarization, and a total expected integrated luminosity of $L=5$~ab$^{-1}$. To disentangle the intrinsic performance of the recoil-mass method from collider-specific effects, two signal configurations are considered throughout this study: an unpolarized sample without ISR and a more realistic scenario including ISR with $P_{e^-}=-80\%$.  

Signal and SM background samples are generated at leading order with \textsc{MadGraph5\_\allowbreak aMC@NLO} \cite{Alwall:2014}. To take spin correlations into account, we use \textsc{MadSpin}~\cite{Artoisenet:2013} to simulate the decay chain, and further perform the parton showering and fragmentation processes with \textsc{Pythia 8}~\cite{Sjostrand:2015}. The final irreducible SM background prediction includes $t\bar{t}$, $t\bar{t}Z$, $t\bar{t}H$, and $W^{+}W^{-}Z$. Additional candidate backgrounds, $t\bar{t}\nu_e\bar{\nu}_e$, $H\nu_e\bar{\nu}_e$, $W^{+}W^{-}$, $ZZ$, and $HZ$, were also generated. The latter are completely removed by the selection and are therefore discarded from the final background estimate. For simplicity, we neglect reducible backgrounds that arise from particles reaching the hadronic calorimeters of the detector.

The analysis targets benchmark top partner masses $m_{T}$ between 1.2 and 2.4~TeV, in 400~GeV steps, with widths derived from the simulation campaign. Tables~\ref{tab:Tt1200Kappa}--\ref{tab:Tt2400Kappa} (see App.~\ref{app:widths}) summarize the $\kappa_{T}$-dependent widths extracted from the simulations.~Unless otherwise stated, the coupling strength is set to $\kappa_{T}=0.2$ for the baseline sensitivity studies, corresponding to a total width of 22.3~GeV for $m_{T} = 1.2$~TeV.

\section{Event selection and reconstruction}

The analysis strategy consists of reconstructing the hadronic decay of a SM top quark as a large radius (large-$R$) jet and calculating its recoil mass. All reconstructed input objects are restricted to the fiducial region $|\eta|<2.5$ before jet clustering, corresponding to the central detector acceptance in which the reconstruction of jets and their substructure is reliable. The event selection then demands at least one Cambridge/Aachen\cite{Dokshitzer_1997,wobisch1999hadronizationcorrectionsjetcross} large-$R$ jet with radius $R=1.0$ and $p_{T}>100$~GeV. This large-$R$ jet must be tagged with the Johns Hopkins top-tagger~\cite{Kaplan:2008} and at least one of its resolved subjets must be $b$-tagged. The tagger is applied to the large-$R$ jet substructure to identify a hard three-prong topology compatible with a boosted hadronic top decay, while the $b$-tag requirement suppresses light-flavour QCD jets and EW backgrounds that can otherwise pass the recoil-mass selection.

The recoil mass is computed event-by-event by solving
\begin{equation}
  m_{\text{recoil}}^{2} = s - 2\,E_{\text{top}}\sqrt{s} + m_{\text{top}}^{2},
  \label{eq:recoilMass}
\end{equation}
where $E_{\text{top}}$ and $m_{\text{top}}$ are reconstructed from the leading large-$R$ jet that passes the object selection requirements. Further selection criteria are applied to reduce the SM background contamination, namely,
\begin{itemize}
  \item total missing energy in the center-of-mass frame, $E_{\text{miss}} < 1$~TeV, which primarily suppresses the $W^{+}W^{-}Z$ and $t\bar{t}Z$ processes, and
  \item $m_{\text{recoil}} > 1.1$~TeV, to suppress the SM continuum.
\end{itemize}

These are the only requirements applied in order to maintain a decay-mode and mass-agnostic analysis. As $m_{T}$ grows larger, the top partner becomes less boosted due to phase-space suppression, so any additional requirements on the event kinematics would bias the reconstructed recoil-mass spectra.

\begin{figure}[!ht]
  \centering
  \includegraphics[width=0.65\textwidth]{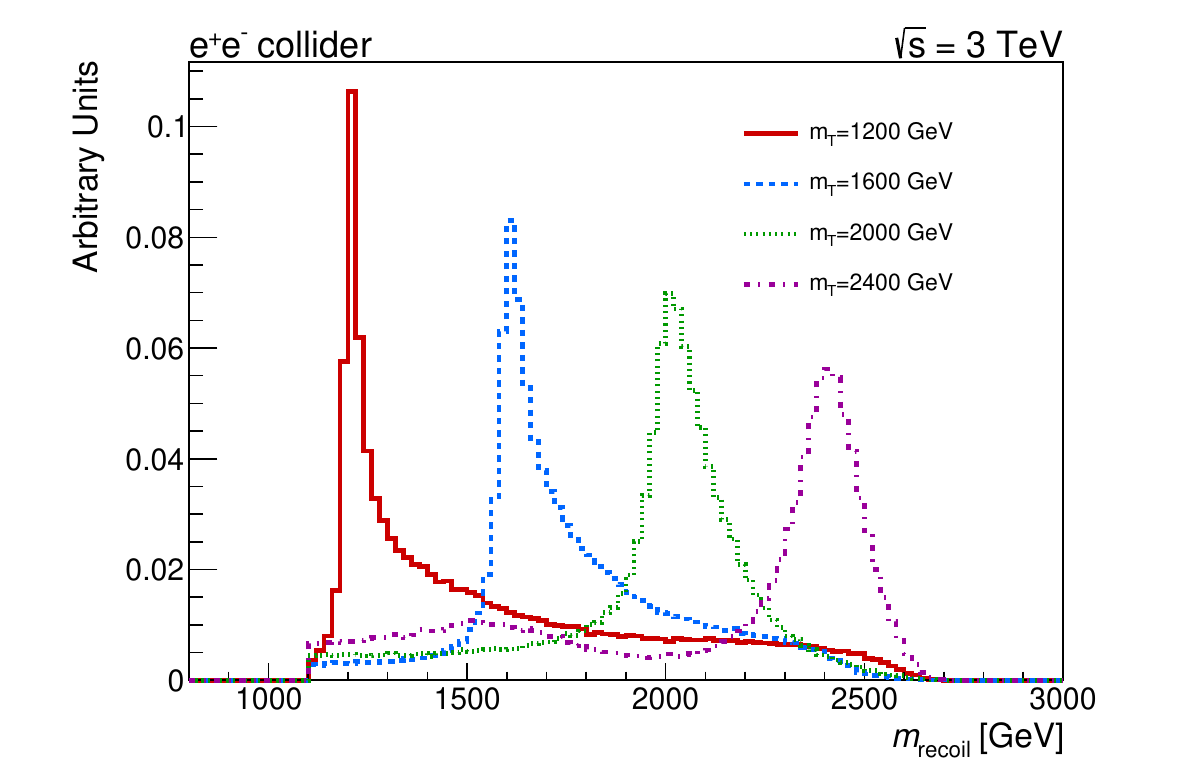}
  \caption{Recoil mass of the signal samples after the baseline selection cuts.}
  \label{fig:signal_after_cut}
\end{figure}

The reconstructed recoil-mass distributions for the different signal samples are shown in Figure~\ref{fig:signal_after_cut}. A clear peak is observed around the nominal mass values for all samples. For the larger masses, a broad tail develops at the low end of the spectra. Generator-level studies indicate that this feature is associated with selecting the secondary top quark originating from the $T$ decay chain rather than the associated top quark produced in $e^{+}e^{-}\to Tt$.

\section{Idealized baseline calibration}

Table~\ref{tab:baseline-cutflow} lists the weighted event yields for $m_{T}=1.2$~TeV together with the dominant $t\bar{t}$ background after the main selection steps for the unpolarized, no-ISR samples.

\begin{table}[!ht]
  \centering
  \caption{Baseline cutflow yields for the $m_{T}=1.2$~TeV signal with $\kappa_T = 0.2$ and for the $t\bar{t}$ background, scaled to $\SI{5}{ab^{-1}}$.}
  \label{tab:baseline-cutflow}
  \begin{tabular}{l S[table-format=5.1] S[table-format=6.1]}
    \toprule
    Selection step & {Signal} & {$t\bar{t}$} \\
    \midrule
    All events                & 1232.3 & 95707.8 \\
    Large-$R$ jet preselection      & 1232.0 & 95338.5 \\
    Top tagged                & 401.4  & 41503.8 \\
    Substructure tag          & 348.6  & 37864.5 \\
    Isolation veto            & 215.1  & 31983.7 \\
    $m_{\text{recoil}}>\SI{1.1}{TeV}$ & 206.1  & 5922.1 \\
    $E_{\text{miss}}<\SI{1}{TeV}$     & 198.6  & 5680.7 \\
    \bottomrule
  \end{tabular}
\end{table}

Due to the potential presence of additional SM top quarks arising from the $T$ decay chain, an isolation requirement is applied to the large-$R$ jets, vetoing candidates with nearby activity either from high-$p_{T}$ leptons or small-radius jets. In practice, the candidate is rejected when either an isolated charged lepton or an additional small-$R$ jet is nearby, as this indicates that the reconstructed large-$R$ jet is embedded in a broader top-decay environment rather than representing the associated hadronic top used for the recoil mass measurement. This requirement is imposed only on large-$R$ jets that carry a relative $H_T$\footnote{$H_T$ is defined as the scalar sum of the $p_{T}$ of all small-$R$ ($R=0.5$) Cambridge/Aachen jets in the event with $p_{T} > 20$~GeV. The relative $H_T$ is the fraction of the total $H_T$ carried by the selected large-$R$ jet.} above 0.4. The veto therefore suppresses candidates with a large share of the event $H_T$ when additional resolved activity is present, reducing semileptonic SM top-quark backgrounds that would otherwise mimic the boosted-top signature.

\begin{figure}[!ht]
  \centering
  \includegraphics[width=0.65\textwidth]{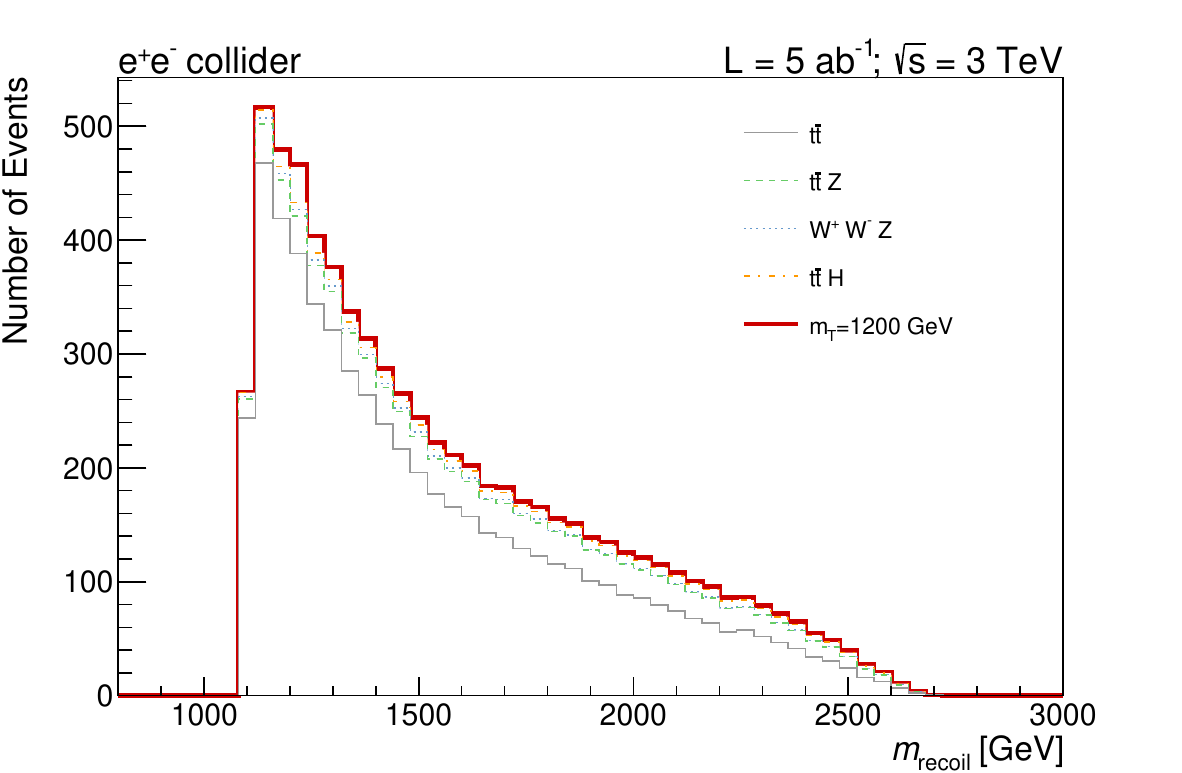}
  \caption{Baseline recoil mass spectrum for $\kappa_T=0.2$ with no polarization or ISR.  Histograms are normalized to $5$~ab$^{-1}$.}
  \label{fig:baseline-recoil}
\end{figure}

After the missing-energy requirement, the $t\bar{t}$ sample is reduced to about \num{5.7e3} events while the signal retains \num{1.99e2} events. Including the subleading EW contributions brings the total background expectation to \num{6.9e3} events. The corresponding signal yields and counting significances, defined as $Z = S/\sqrt{S + B}$, are summarized in Table~\ref{tab:significance-summary} for each benchmark mass point.  Figure~\ref{fig:baseline-recoil} displays the recoil mass spectrum for the $m_{T}=1.2$~TeV signal stacked on top of the total SM background expectation.

\section{Realistic discovery reach}

\begin{figure}[!ht]
  \centering
  \includegraphics[width=0.47\textwidth]{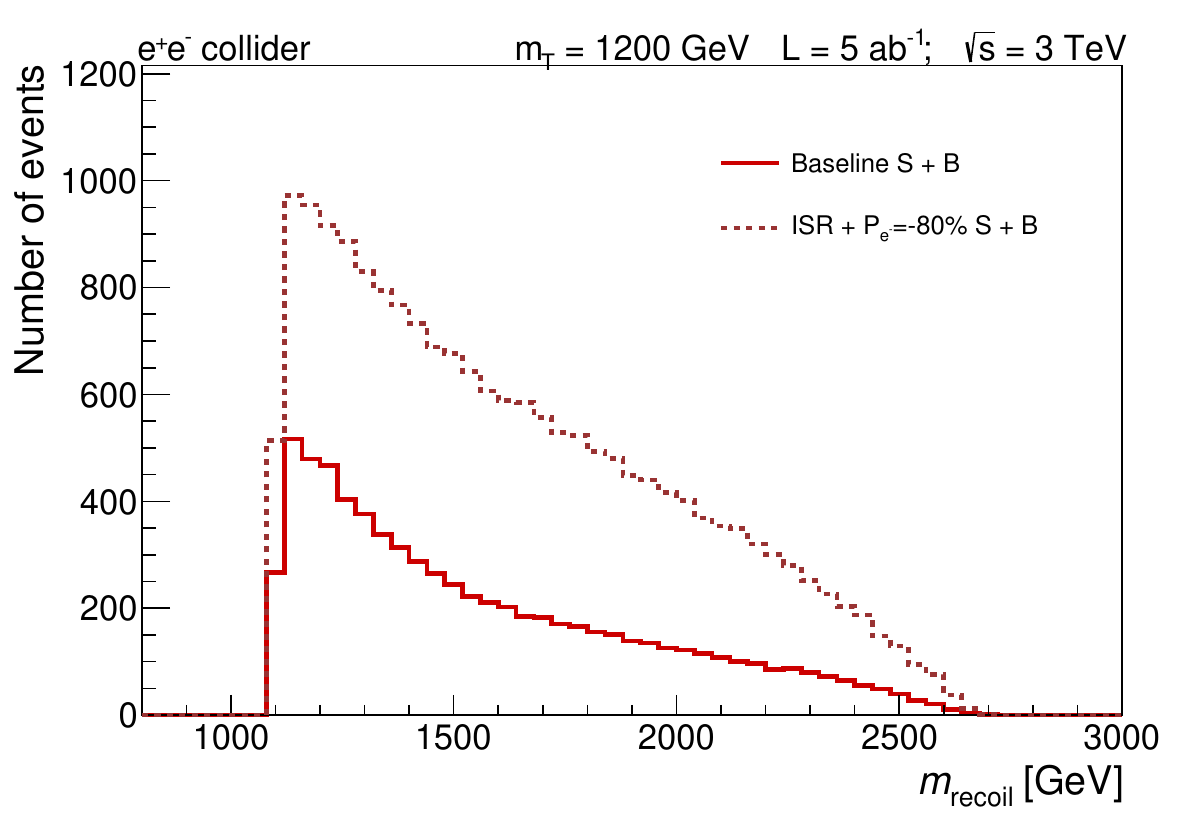}\hfill
  \includegraphics[width=0.47\textwidth]{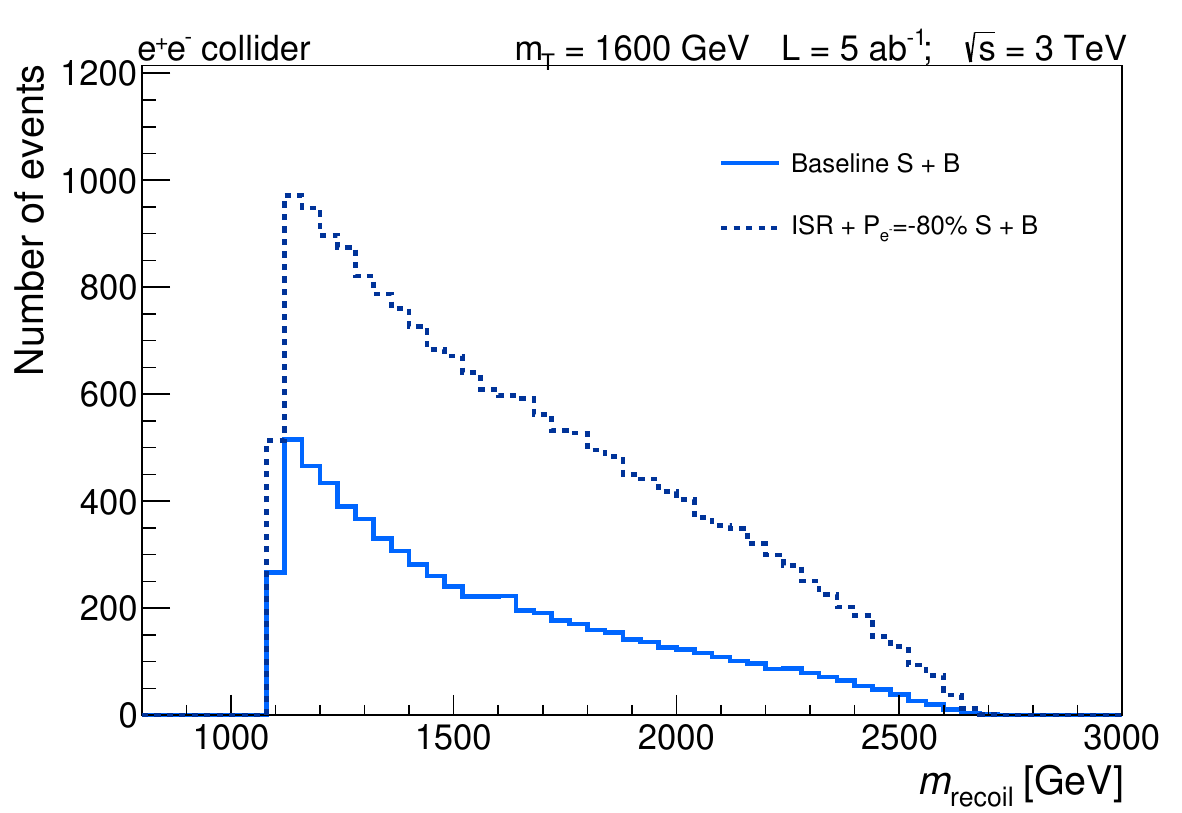}

  \vspace{0.5em}

  \includegraphics[width=0.47\textwidth]{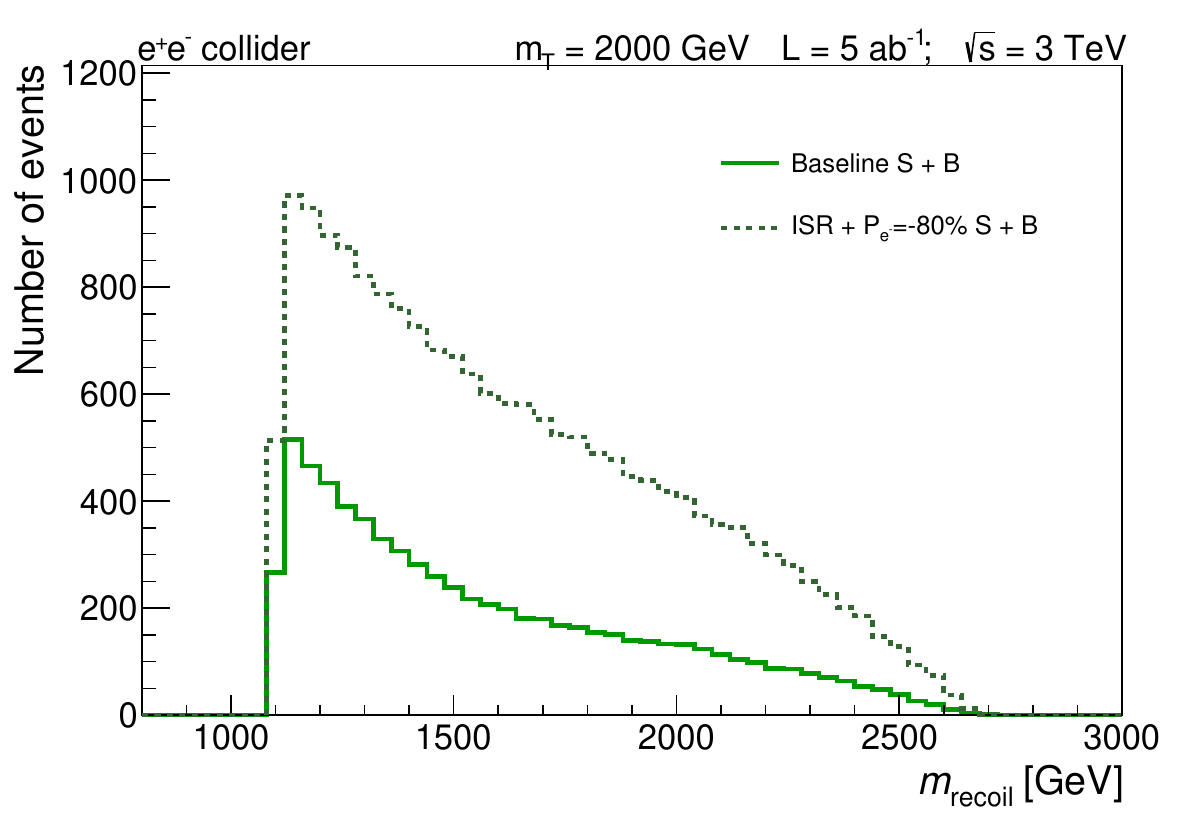}\hfill
  \includegraphics[width=0.47\textwidth]{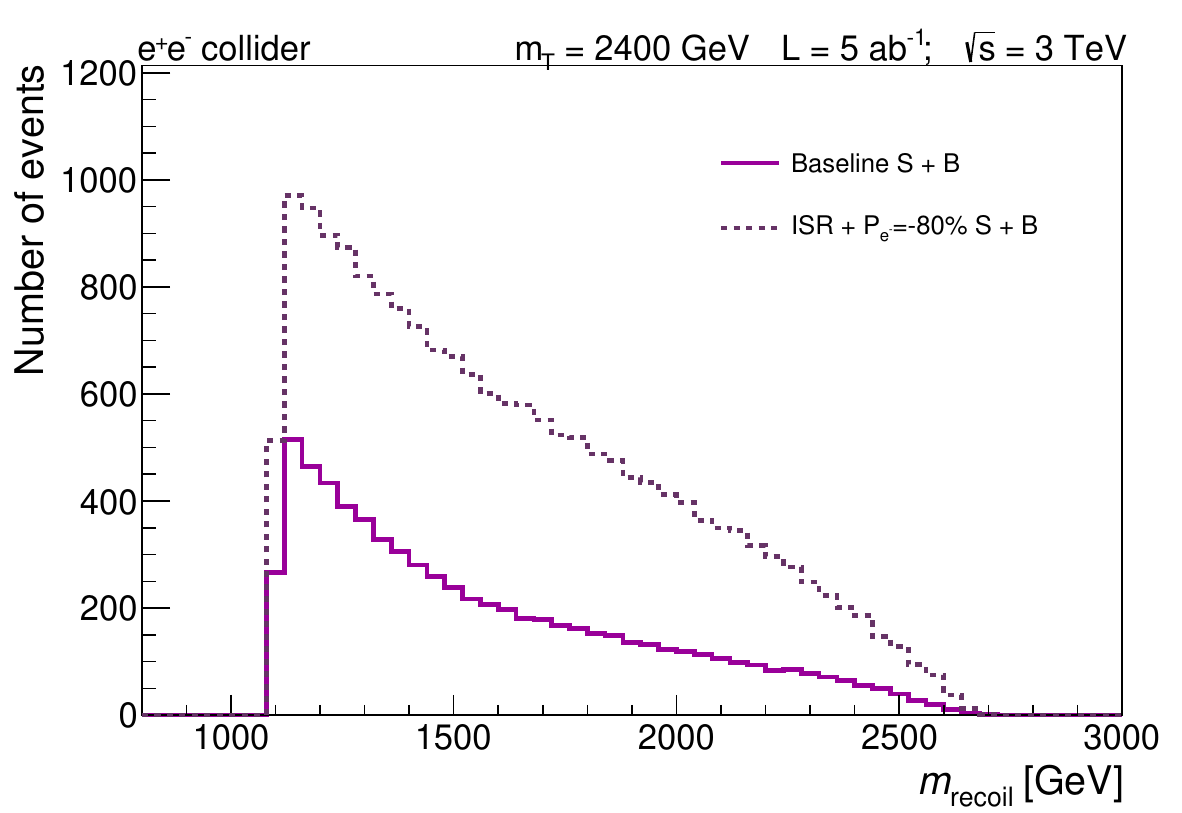}

  \caption{Total signal-plus-background recoil-mass spectra for the four mass benchmarks. Reading order: 1200, 1600, 2000, and 2400 GeV. Each panel compares the idealized unpolarized baseline without ISR to the ISR plus electron-polarization configuration with $P_{e^-}=-80\%$.}
  \label{fig:comparison-recoil}
\end{figure}

To assess the discovery reach under conditions closer to those expected during CLIC operation, the analysis is repeated on samples that include ISR together with an electron-beam polarization of $-80\%$ (positrons remain unpolarized). Because the recoil-mass technique relies on the nominal center-of-mass energy, ISR broadens the reconstructed peak, while the change in beam polarization modifies both the signal rate and the background composition. Figure~\ref{fig:comparison-recoil} compares the total signal-plus-background expectation in the two configurations. The significance loss is summarized in Table~\ref{tab:significance-summary}; for $m_{T}=1.2$~TeV the discovery reach drops from $Z=2.35$ to $Z=1.54$.

The observed loss in sensitivity also points to concrete mitigation handles for future analyses. ISR photons are emitted predominantly along the beam direction and therefore remove longitudinal energy while leaving the transverse structure of the event comparatively stable. CLIC recoil-mass studies that include ISR and \textit{beam-strahlung} have used transverse-momentum imbalance and acolinearity as discriminating inputs~\cite{CLICHiggsISR}. This motivates investigating the corresponding recoil-system transverse-momentum imbalance and top--recoil acolinearity in a multivariate discriminant for the present channel, with the aim of recovering part of the lost significance while preserving the mass- and decay-agnostic character of the search.

\begin{table}[!ht]
  \centering
  \caption{Signal yields ($S$) and counting significances ($Z$) at $L=\SI{5}{ab^{-1}}$ for the unpolarized baseline without ISR and the ISR+$P_{e^{-}}=-80\%$ samples.}
  \label{tab:significance-summary}
  \begin{tabular}{c S[table-format=3.1] c S[table-format=1.2] c S[table-format=3.1] c S[table-format=1.2]}
    \toprule
    $m_{T}$ [TeV] & {$S_{\text{base}}$} & & {$Z_{\text{base}}$} & & {$S_{\text{ISR},-80}$} & &{$Z_{\text{ISR},-80}$} \\
    \midrule
    1.2 & 198.9 & & 2.35 & & 211.7 & & 1.54 \\
    1.6 & 165.6 & & 1.96 & & 160.5 & & 1.17 \\
    2.0 & 97.0  & & 1.16 & & 81.6 & & 0.60 \\
    2.4 & 25.6  & & 0.31 & & 18.0 & & 0.13 \\
    \bottomrule
  \end{tabular}
\end{table}

\section{Scanning $\kappa_T$}
\begin{figure}[!ht]
  \centering
  \includegraphics[width=0.48\textwidth]{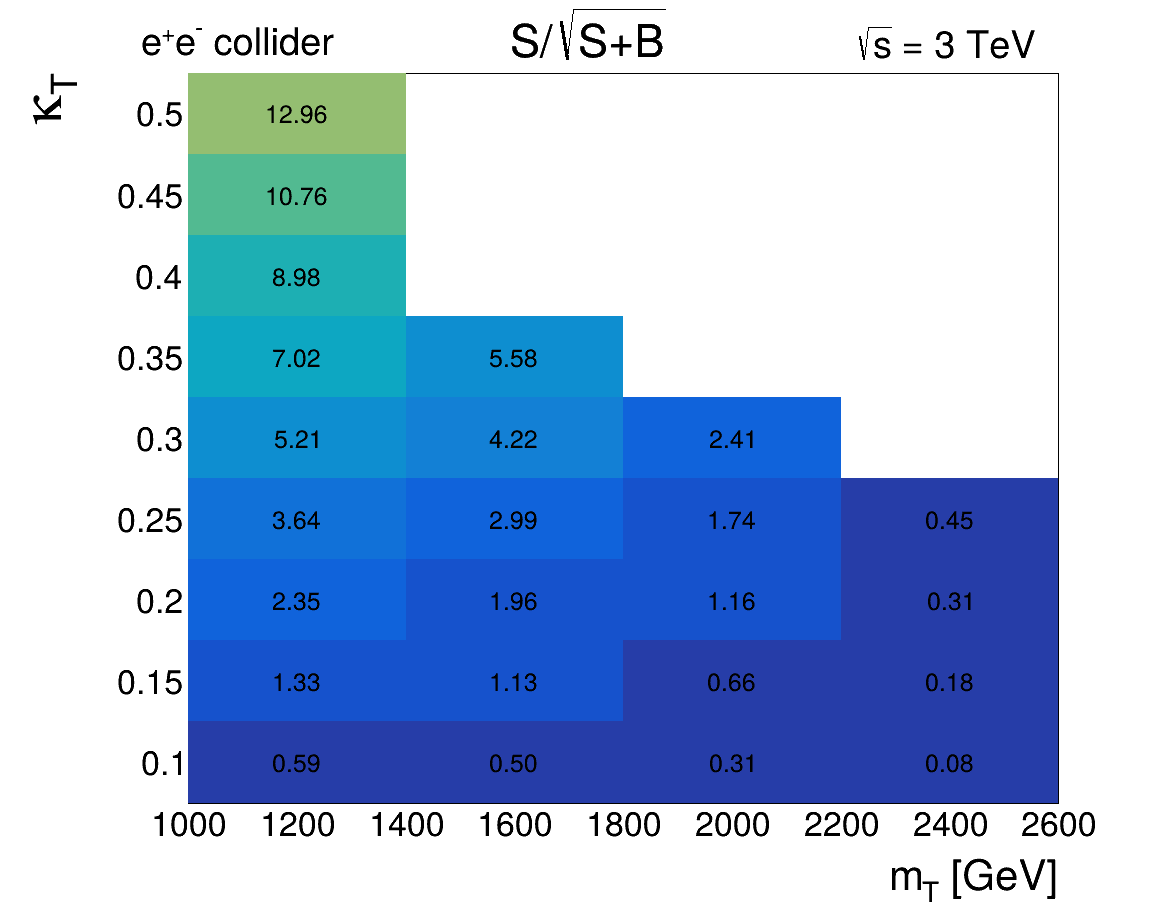}\hfill
  \includegraphics[width=0.48\textwidth]{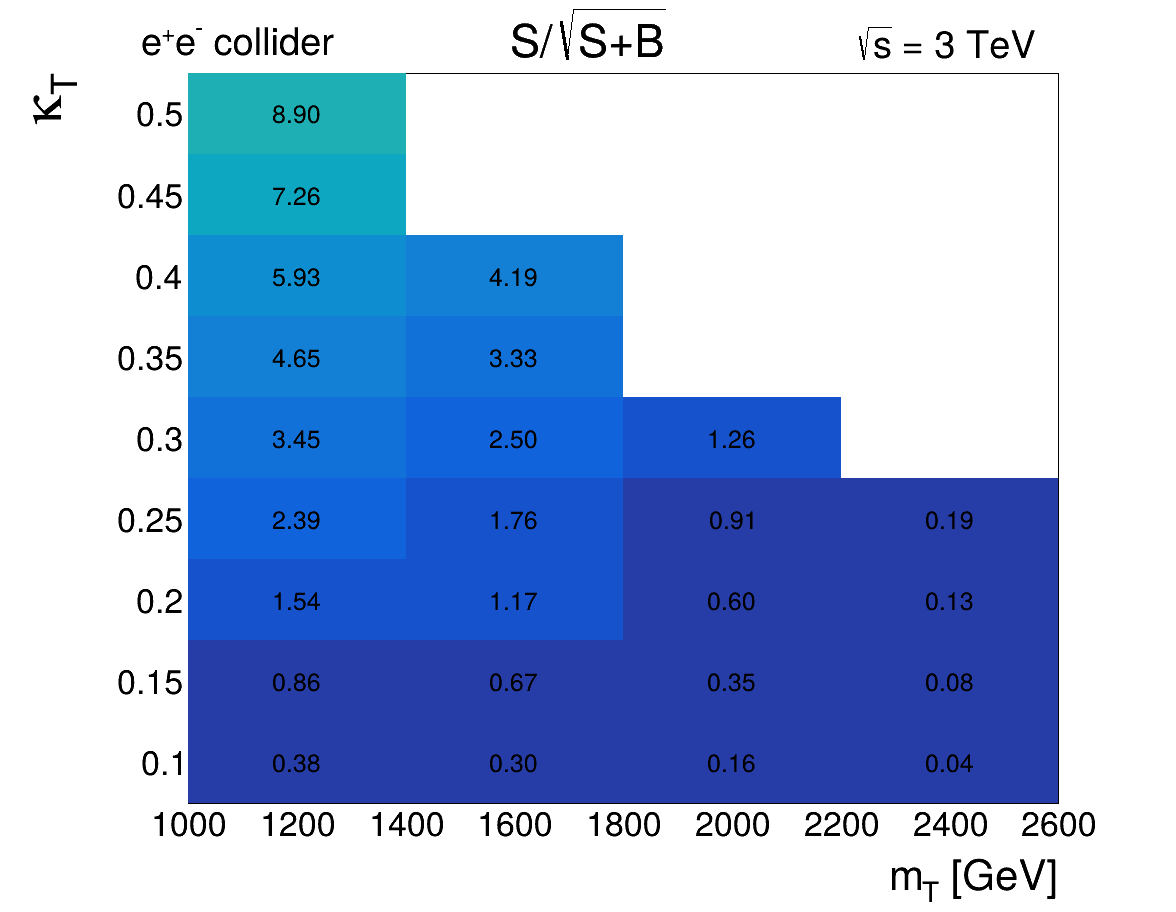}
  \caption{Significance across the mass-coupling plane. Left: idealized baseline without ISR or beam polarization. Right: ISR plus electron polarization with $P_{e^-}=-80\%$.}
  \label{fig:kappa-scans}
\end{figure}

To evaluate the sensitivity of the analysis strategy, the counting significance is estimated for different $\kappa_T$ values. The resulting maps in the $(m_{T},\kappa_T)$ plane are visualized side by side in Fig.~\ref{fig:kappa-scans}, comparing the idealized baseline to the ISR+$P_{e^-}=-80\%$ configuration. The baseline configuration reaches $Z\simeq 5$ for $(m_{T},\kappa_T)=(1.2$~TeV$,0.3)$. At $m_{T}=1.2$~TeV, the baseline already exceeds $Z=1$ at $\kappa_T=0.2$, whereas the ISR+$P_{e^-}=-80\%$ configuration reaches $Z>1$ near $\kappa_T=0.3$; at $m_{T}=2.4$~TeV, the reported points remain below unity. The sections of the $(m_T, \kappa_T)$ plane with no reported value (white) correspond to the configurations in which the simulations yielded unphysical results (e.g. extremely large widths).

These contours provide a transparent summary that can be reinterpreted for specific decay modes by rescaling the signal efficiencies.

\section{Discussion and outlook}
A prospective study of single production of a top partner $T$ at a $\sqrt{s}=3$~TeV electron-positron collider has been presented using an analysis strategy that does not depend on the assumed mass or decay mode. The recoil-based selection remains robust against variations in the top-partner mass and already yields $\mathcal{O}(2\sigma)$ sensitivity at $\kappa_T=0.2$ in the idealized baseline configuration. The comparison with the ISR+$P_{e^-}=-80\%$ scenario then makes explicit how realistic collider effects reduce the reach by roughly $35\%$ in significance, while simultaneously identifying the global observables that should be prioritized to mitigate this loss.

A benchmark analysis setup is presented for future lepton-collider searches in the regions where channel-specific hadron-collider strategies deteriorate as branching-fraction assumptions fail. Further optimization of ISR-sensitive observables, together with detector and reconstruction improvements to the hadronic and substructure resolution of boosted large-$R$ jets, can strengthen the decay-agnostic recoil technique. More broadly, these results show that a mass- and decay-agnostic recoil strategy can become a useful tool for probing vector-like top partners with minimal model dependence across the $(m_T,\kappa_T)$ parameter space.

\section*{Data Availability Statement}
The analysis code and configuration files are publicly available at \url{https://github.com/zadermugiwara/ttp_Analysis/tree/main}. The simulated event samples used in this study are not publicly available because of their size and the absence of a suitable public repository for hosting them. Additional numerical outputs are available from the corresponding author upon reasonable request.

\section*{AI Disclosure}

OpenAI Codex (GPT-5, OpenAI; accessed August--September 2026) was used to assist with language editing, manuscript organization, \LaTeX{} formatting, the identification of potentially relevant literature, and the preparation of documentation for the analysis code referenced in the Data Availability Statement. The authors directed its use through targeted prompts and independently reviewed all generated text and code documentation, consulted the original literature, and verified the scientific statements and numerical results against the original analysis outputs. The AI tool was not used to generate the simulated event samples, perform the data analysis or statistical calculations, or produce the scientific results presented in this work. The authors take full responsibility for the content of the manuscript and the accuracy of the shared code documentation.

\section*{Acknowledgements}

This work was supported by the Dirección de Investigación y Posgrado (DINVP) of the Universidad Iberoamericana Ciudad de México.

\appendix
\section{Width tables}\label{app:widths}
\begin{table}[!ht]
\centering
\caption{Widths $\Gamma(T)$ for the $m_T=1.2$~TeV samples as a function of $\kappa_T$.}
\label{tab:Tt1200Kappa}
\begin{tabular}{lr}
\toprule
$\kappa_T$ & Width [GeV] \\
\midrule
$0.10$ & 5.58 \\
$0.15$ & 12.55 \\
$0.20$ & 22.32 \\
$0.25$ & 34.87 \\
$0.30$ & 50.22 \\
$0.35$ & 68.35 \\
$0.40$ & 89.27 \\
$0.45$ & 112.98 \\
$0.50$ & 139.49 \\
\bottomrule
\end{tabular}
\end{table}

\begin{table}[!ht]
\centering
\caption{Widths $\Gamma(T)$ for the $m_T=1.6$~TeV samples as a function of $\kappa_T$.}
\label{tab:Tt1600Kappa}
\begin{tabular}{lr}
\toprule
$\kappa_T$ & Width [GeV] \\
\midrule
$0.10$ & 13.32 \\
$0.15$ & 29.96 \\
$0.20$ & 53.27 \\
$0.25$ & 83.24 \\
$0.30$ & 119.86 \\
$0.35$ & 163.14 \\
$0.40$ & 213.08 \\
\bottomrule
\end{tabular}
\end{table}

\begin{table}[!ht]
\centering
\caption{Widths $\Gamma(T)$ for the $m_T=2$~TeV samples as a function of $\kappa_T$.}
\label{tab:Tt2000Kappa}
\begin{tabular}{lr}
\toprule
$\kappa_T$ & Width [GeV] \\
\midrule
$0.10$ & 26.09 \\
$0.15$ & 58.71 \\
$0.20$ & 104.38 \\
$0.25$ & 163.09 \\
$0.30$ & 234.86 \\
\bottomrule
\end{tabular}
\end{table}

\clearpage
\begin{longtable}{lr}
\caption{Widths $\Gamma(T)$ for the $m_T=2.4$~TeV samples as a function of $\kappa_T$.}
\label{tab:Tt2400Kappa} \\
\toprule
$\kappa_T$ & Width [GeV] \\
\midrule
$0.10$ & 45.17 \\
$0.15$ & 101.63 \\
$0.20$ & 180.68 \\
$0.25$ & 282.31 \\
\bottomrule
\end{longtable}
\clearpage

\bibliographystyle{apsrev4-2}
\bibliography{references}

\end{document}